\documentclass[a4paper]{PoS}

\usepackage{lineno}
\usepackage[subrefformat=parens,labelformat=parens]{subfig}
\usepackage{xspace}
\usepackage{cite}
\usepackage[utf8]{inputenc}
\usepackage{floatrow}
\usepackage[all]{nowidow}
\usepackage{multicol}

\newcommand{\Subref}[1]{\protect\subref{#1}} 

\def\lc{lightcurve\xspace}
\def\lcs{lightcurves\xspace}
\def\zdcf{$Z$DCF\xspace}
\def\syncFreq{\ensuremath{\nu_{\mathrm{sync}}}\xspace}

\def\veritas{VERITAS\xspace}
\def\lat{\textit{Fermi}-LAT\xspace}
\def\xrt{\textit{Swift}-XRT\xspace}

\def\esZero{1ES~0033+595\xspace}
\def\esFive{1ES~0502+675\xspace}
\def\esTen{1ES~1011+496\xspace}
\def\esTwelve{1ES~1218+304\xspace}
\def\esTwo{1ES~0229+200\xspace}
\def\rgbj{RGB~J0710+591\xspace}
\def\pg{PG~1553+113\xspace}

\newcommand*{\TeV}{\ifmmode {\mathrm{\ Te\kern -0.1em V}\xspace}\else
                   \textrm{Te\kern -0.1em V}\xspace\fi}%
\newcommand*{\GeV}{\ifmmode {\mathrm{\ Ge\kern -0.1em V}\xspace}\else
                   \textrm{Ge\kern -0.1em V}\xspace\fi}%
\newcommand*{\MeV}{\ifmmode {\mathrm{\ Me\kern -0.1em V}\xspace}\else
                   \textrm{Me\kern -0.1em V}\xspace\fi}%
\newcommand*{\keV}{\ifmmode {\mathrm{\ ke\kern -0.1em V}\xspace}\else
                   \textrm{ke\kern -0.1em V}\xspace\fi}%
\newcommand*{\eV}{\ifmmode  {\mathrm{\ e\kern -0.1em V}\xspace}\else
                   \textrm{e\kern -0.1em V}\xspace\fi}%
\let\tev=\TeV
\let\gev=\GeV
\let\mev=\MeV
\let\kev=\keV
\let\ev=\eV

\DeclareFloatVCode{afterfloat}{\vspace{-15pt}}
\DeclareFloatVCode{beforefloat}{\vspace{-15pt}}

\title{Variability Study of Extreme Blazars with \veritas}

\ShortTitle{Variability Study of Extreme Blazars with \veritas}

\author{\speaker{Orel Gueta}, for the \veritas collaboration\footnote{https://veritas.sao.arizona.edu/} \footnote{for collaboration list see PoS(ICRC2019)1177}\\
        Deutsches Elektronen-Synchrotron (DESY), Zeuthen, Germany\\
        E-mail: \email{orel.gueta@desy.de}}

\abstract{The \veritas array of imaging atmospheric Cherenkov telescopes has collected nearly 6000 hours of observations of active galactic nuclei (AGNs). It has detected 39 very-high-energy (VHE, >100 \gev) AGNs at redshifts up to $z = 0.9$, of which 24 are classified as high-frequency-peaked BL Lacertae objects (HBLs). \veritas has obtained an extensive dataset of HBL, including extreme HBL (xHBL), observations, with \lcs spanning up to 11 years, allowing the characterization of their long-term and short-term variability. A study of various xHBLs/HBLs in the \veritas dataset is presented, and the correlation with other energy bands tested. In particular, the short-term variability of xHBLs as a function of energy within the VHE band is examined, exploring the possibility that secondary gamma rays are produced in cosmic-ray interactions with background photons.}

\FullConference{36th International Cosmic Ray Conference -ICRC2019-\\
		July 24th - August 1st, 2019\\
		Madison, WI, U.S.A.}

\begin{document}

\section{introduction}

\noindent The blazar population, a class of active galactic nuclei (AGNs), dominates the extragalactic gamma-ray sky. 
Of those, the most common are BL Lac type blazars~\cite{Fermi-LAT:2019yla}.
Their spectral energy distribution (SED) typically features two main broad peaks; a low energy one commonly explained as synchrotron emission by highly relativistic electrons in the jet, and a higher energy one typically attributed to inverse-Compton emission of the same electrons off seed photons.
The seed photons can be synchrotron photons themselves (synchrotron self-Compton, SSC)~\cite{1985ApJ...298..114M} or those originating from outside the jet (e.g., from the accretion disk)~\cite{1993ApJ...416..458D}.
BL Lac type blazars are further classified according to the frequency of their synchrotron peak (\syncFreq) into high-frequency-peaked (HBLs, $\syncFreq > 10^{15}\mathrm{Hz}$), intermediate-frequency-peaked (IBL, $10^{14}\mathrm{Hz} < \syncFreq < 10^{15}\mathrm{Hz}$) and low-frequency-peaked BL Lac objects (LBLs, $\syncFreq < 10^{14}\mathrm{Hz}$)~\cite{Padovani:1994sh}.
HBLs currently account for the largest fraction of \tev detected blazars, with 51 HBLs detected so far.\footnote{http://tevcat.uchicago.edu/}
The majority of the sources studied here belong to a subset of HBLs characterized by synchrotron peak frequencies $\gtrsim 10^{17}$~Hz, and referred to as extreme HBLs (xHBLs)~\cite{Senturk:2013pa}.
The observed SEDs of these xHBLs pose challenges to the conventional one-zone SSC model, which has been relatively successful in describing less extreme HBL spectra~\cite{Ghisellini:1998it}.
Several mechanisms within the SSC picture were suggested to model the SEDs of xHBLs and in particular to explain their observed \tev spectra~\cite{Lefa_2011}.
An alternative solution for the latter was put forth in Ref.~\cite{Essey:2009zg}, proposing that the gamma ray flux from distant blazars is dominated by secondary gamma rays. 
Those are produced along the line of sight by the interactions of protons originating from the blazar with background photons.
Studying the time variability of these xHBLs/HBLs and searching for correlations between different wavelengths and within the very-high-energy band (VHE, $> 100$ \gev) could help constrain or reject some of these models.
For this purpose, the \lcs of \esZero, \esFive, \esTen, \esTwelve, \esTwo, \rgbj and \pg are studied.
The results for \esTen, \esTwelve and \pg are shown in this proceeding.

The \textit{Fermi} large area telescope (LAT), on board the \textit{Fermi} gamma-ray space telescope, is sensitive in the energy range $\sim$20 \mev to $>500$ \gev with a field of view of approximately 2.4 sr~\cite{2009ApJ...697.1071A}.
\lat data are analysed with \textit{Fermipy}~\cite{2017ICRC...35..824W}, using the \texttt{P8R2\_SOURCE\_V6} instrument response function, and the preliminary LAT 8-year Point Source List (FL8Y).

The X-ray Telescope (\xrt) on the Neil Gehrels \textit{Swift} observatory is sensitive to photons with energies between 0.2 and 10 \kev~\cite{Burrows2005}. 
The X-ray \lcs used in this analysis are obtained from the online \xrt \lc repository~\cite{Evans:2007na}.

\veritas~\cite{Park:2015ysa} is located at the Fred Lawrence Whipple Observatory in southern Arizona (31 40N, 110 57W,  1.3~km a.s.l.).
It is sensitive to gamma rays in the energy range, 100 \gev to $> 30$ \tev. 
It can currently detect an object having 1\% Crab Nebula flux within $\sim$25 hours. 
The typical systematic uncertainty on measured flux is $\sim$20\%.
The \veritas AGN observing strategy consists of regular monitoring of known VHE targets in order to identify flaring episodes~\cite{Benbow:2017buy}. 
These observations are complemented by coordinated observations at lower energy, in order to obtain long-term contemporaneous multiwavelength datasets for a variety of AGNs.
This strategy resulted in an extensive dataset of AGNs, containing \lcs covering up to 11 years, opening up the possibility to study their long-term and short-term variability and correlations to other wavelengths.

\section{Multiwavelength \lcs}

\begin{figure}[htp]
\begin{center}
\includegraphics[trim=0mm 0mm 0mm 0mm,clip,width=0.6\textwidth, page=1]{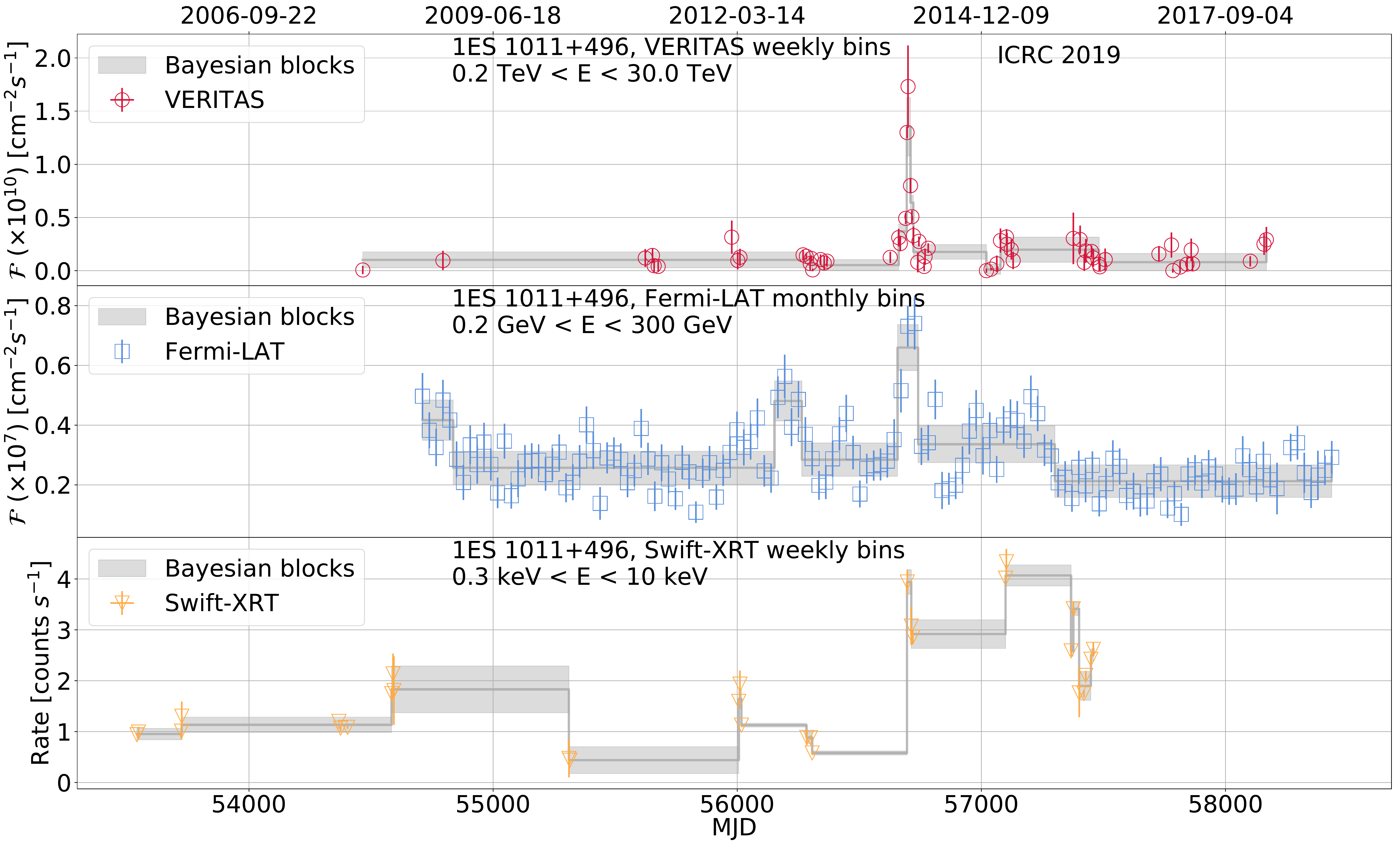}
\caption{\label{fig:1ES1011_weekly}
The \esTen \lcs in VHE (\veritas), HE (\lat) and X-ray (\xrt) in the energy ranges and binning as indicated. The mean flux and the corresponding uncertainty in each Bayesian block are shown in grey.
}
\end{center}
\end{figure}

\begin{figure}[htp]
\begin{center}
\includegraphics[trim=0mm 0mm 0mm 0mm,clip,width=0.6\textwidth, page=1]{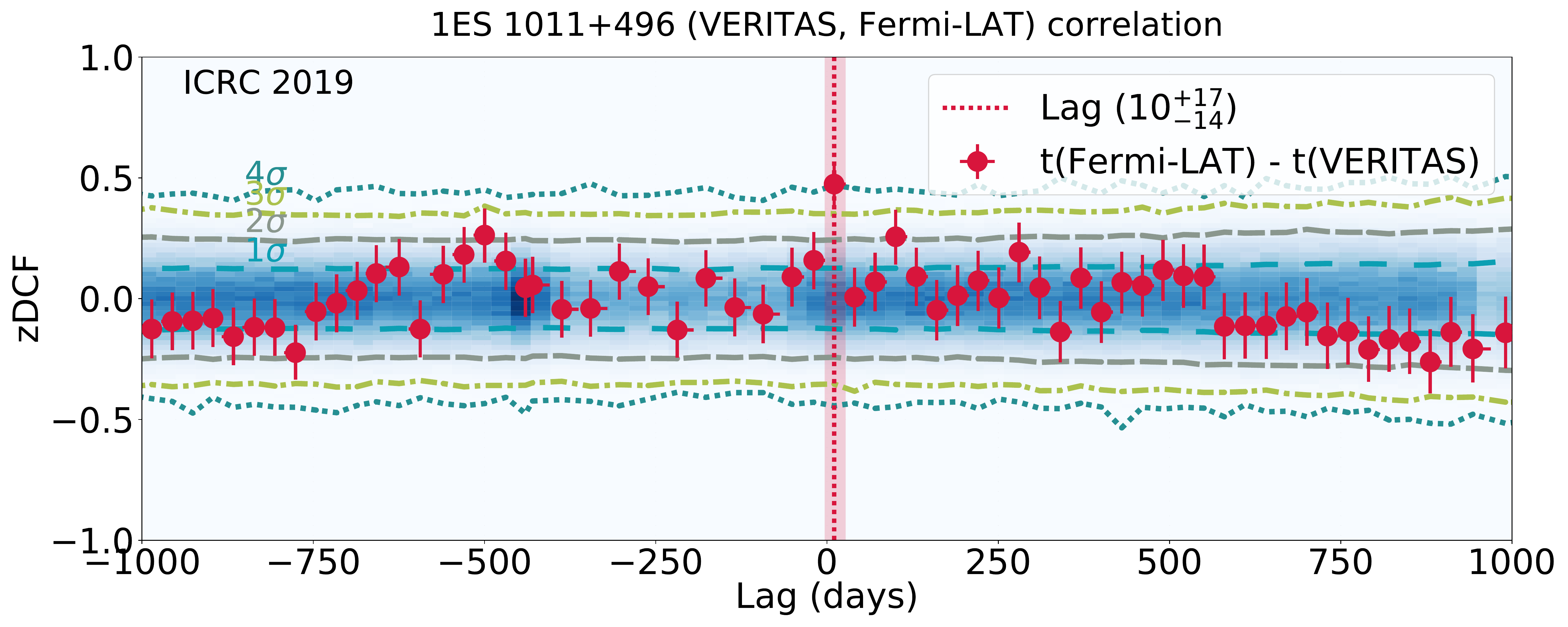}
\caption{\label{fig:1ES1011_zdcf_fermi}
The \zdcf between the VHE (\veritas) and HE (\lat) \esTen \lcs. The vertical red dotted line and band indicate the most likely lag between the two bands and the corresponding 1-$\sigma$ confidence interval. The \zdcf distribution obtained from simulated \lcs is shown in the 2D histogram colour map (see text for details). The correlation significance levels correspond to the quantiles of the \zdcf distribution in each lag bin.
}
\end{center}
\end{figure}

\noindent The weekly \veritas \lc, monthly \lat \lc and weekly \xrt \lc of the \esTen blazar from June 2005 to December 2018 are shown in Figure~\ref{fig:1ES1011_weekly}.
The results of a Bayesian block analysis~\cite{Scargle_2013}, used to determine any significant ($3\sigma$) change points in the \lcs, are shown as well.
A flare during February 2014 is clearly visible in the \veritas and \lat \lcs. 
Hints of a simultaneous high flux state are seen in the \xrt \lc, which seems to extend long after the flare period in \veritas and \lat.
In order to estimate the correlation between the \veritas and \lat \lcs, the $z$-transformed discrete cross-correlation function (\zdcf)~\cite{1997ASSL..218..163A} is calculated.
The \zdcf method provides a conservative estimate for the cross-correlation function as a function of lag between sparsely sampled \lcs. 
The results of the \zdcf algorithm are shown in Figure~\ref{fig:1ES1011_zdcf_fermi}, where evidence for correlation between the \veritas and \lat \lcs is observed with a significance of $4\sigma$. 
The peak time lag is estimated to be $10^{+17}_{-14}$ days, which is consistent with no lag between VHE and high-energy (HE, $0.1 < E < 100$ \gev).
A positive time lag between band $x$ and band $y$, $t(x) - t(y) > 0$, indicates that the emission in band $x$ lags behind band $y$.
The significance of \zdcf values is estimated using the Monte Carlo method.
Simulated \lcs are generated from the power spectral densities (PSD)~\cite{1995A&A...300..707T} of each \lc, and the \zdcf is calculated between them.
The distribution of 100,000 such pseudo-experiments is shown as the blue histogram in Figure~\ref{fig:1ES1011_zdcf_fermi}.
The quantiles of the distribution in each lag value, corresponding to significance levels, are shown as horizontal lines.
The most likely lag value and the corresponding uncertainty are obtained through a maximum likelihood function~\cite{2013arXiv1302.1508A}.
No correlation is observed between the \veritas and \xrt \lcs, probably due to the small amount of \xrt data available, especially during the flare period.
The observed correlation between VHE and HE is expected, as the emission in those energy ranges is typically considered to originate from the same particle population.

\thisfloatsetup{floatwidth=.6\textwidth,capbesidewidth=sidefil,
capposition=beside,capbesideposition={left,center}, postcode=afterfloat}

\thisfloatsetup{captionskip=-10pt}%

\begin{figure}[htp]
\begin{center}
\subfloat[]{\label{fig:1ES1218_nightly}\includegraphics[trim=0mm 0mm 0mm 0mm,clip,width=0.5\textwidth, page=1]{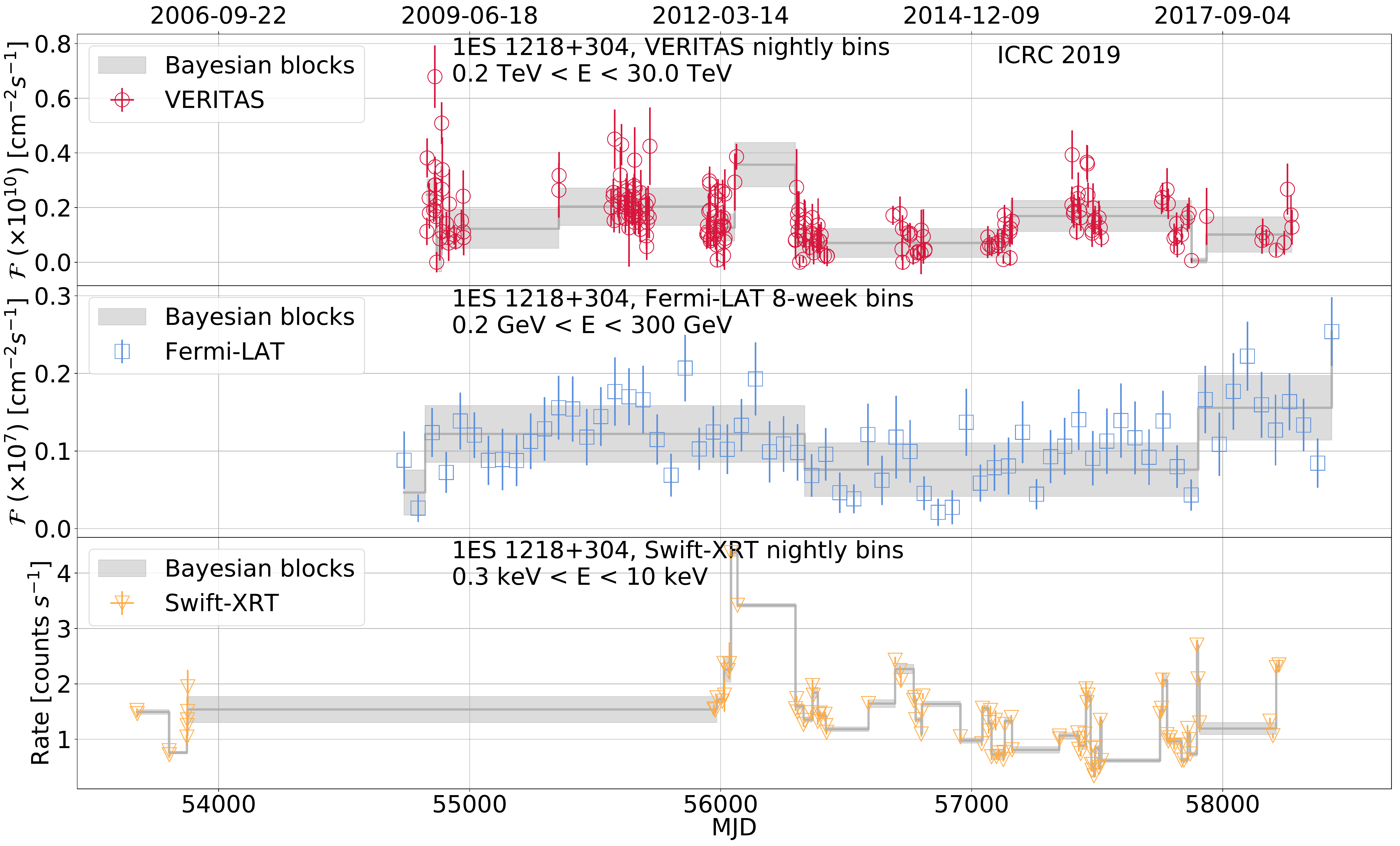}}
\subfloat[]{\label{fig:PG1553_nightly}\includegraphics[trim=0mm 0mm 0mm 0mm,clip,width=0.5\textwidth, page=1]{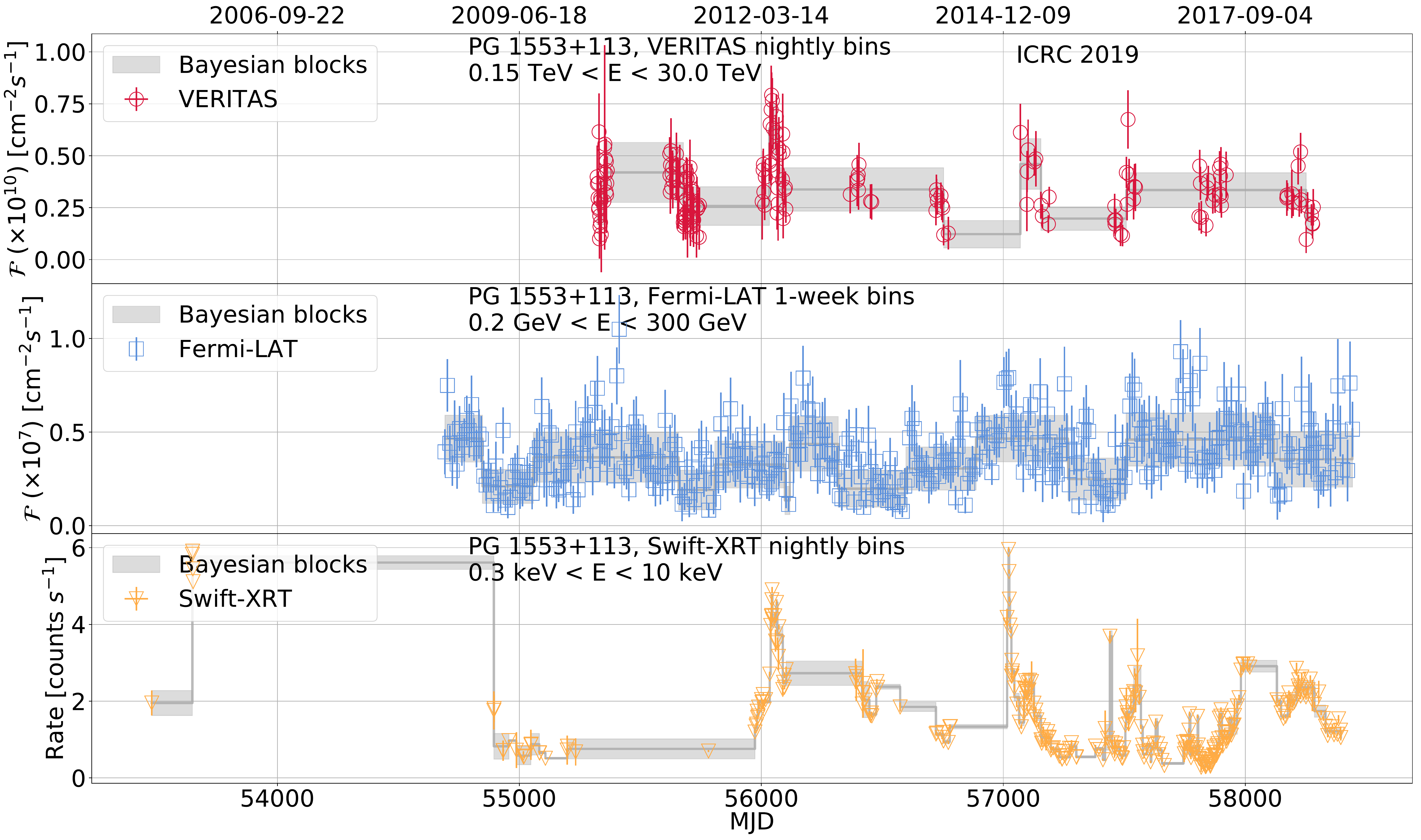}}
\caption{\label{fig:1ES1218_PG1553}
The \Subref{fig:1ES1218_nightly} \esTwelve and \Subref{fig:PG1553_nightly} \pg \lcs in VHE (\veritas), HE (\lat) and X-ray (\xrt) in the energy ranges and binning as indicated. The mean flux and the corresponding uncertainty in each Bayesian block are shown in grey.
}
\end{center}
\end{figure}

\begin{figure}[htp]
\begin{center}
\subfloat[]{\label{fig:1ES1218_zdcf_fermi}\includegraphics[trim=0mm 0mm 0mm 0mm,clip,width=0.5\textwidth, page=1]{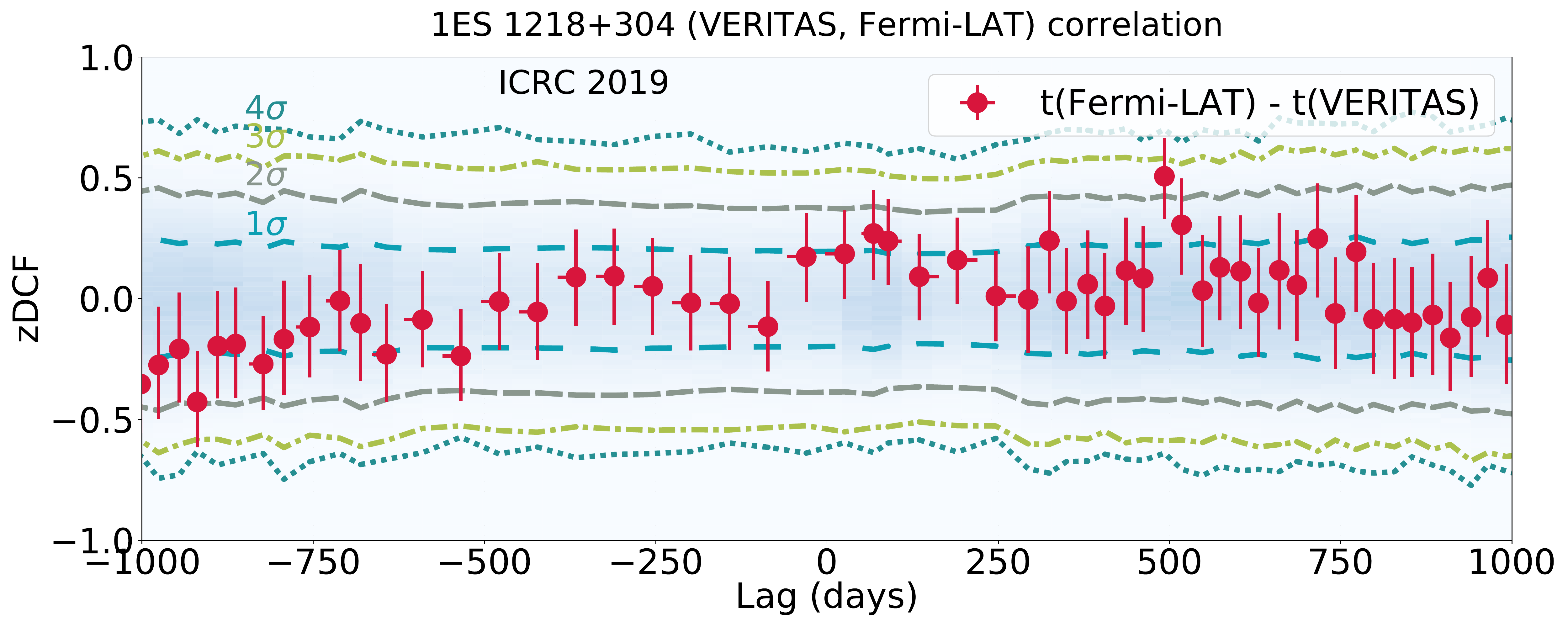}} 
\subfloat[]{\label{fig:1ES1218_zdcf_swift}\includegraphics[trim=0mm 0mm 0mm 0mm,clip,width=0.5\textwidth, page=1]{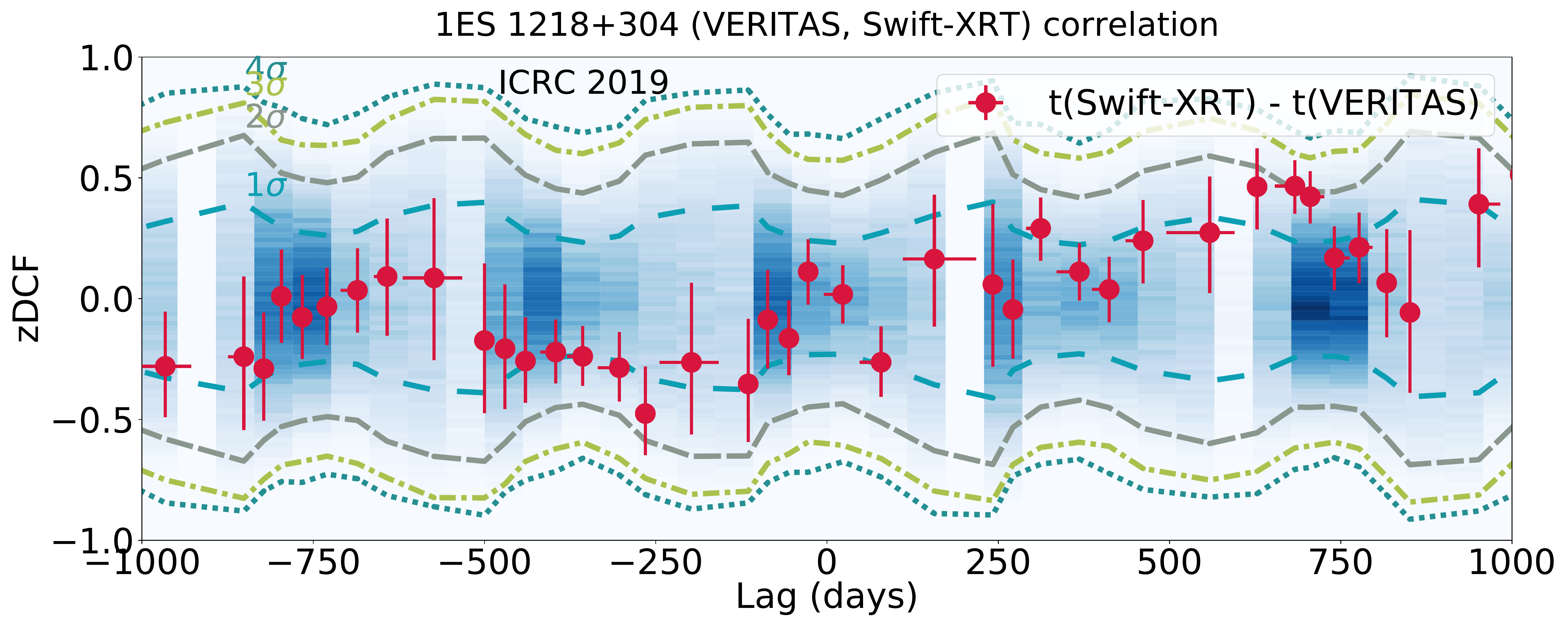}}
\caption{\label{fig:1ES1218_zdcf_mwl}
The \zdcf between the \esTwelve \lcs in \Subref{fig:1ES1218_zdcf_fermi} VHE (\veritas) and HE (\lat); and \Subref{fig:1ES1218_zdcf_swift} VHE (\veritas) and X-ray (\xrt). The \zdcf distribution obtained from simulated \lcs is shown in the 2D histogram colour map (see text for details). The correlation significance levels correspond to the quantiles of the \zdcf distribution in each lag bin.
}
\end{center}
\end{figure}

The nightly \veritas and \xrt \lcs of \esTwelve, and the corresponding \lat \lc, divided to 8-week bins, are shown in Figure~\subref*{fig:1ES1218_nightly}.
Variability is observed in all energy bands, with short time-scale variability seen in X-ray; month-scale variability in VHE; and long-term variability in HE.
The 2009 VHE flare, where \veritas observed intra-night variability~\cite{2010ApJ...709L.163A}, is seen as well.
Figure~\ref{fig:1ES1218_zdcf_mwl} shows the \zdcf results between the \veritas \lc and the \lat and \xrt ones. 
No correlation is observed between any of the energy bands, including HE and X-ray (not shown).
The lack of correlation between VHE and HE could be due to the sensitivity of \lat being insufficient to detect the shorter-term variability observed with \veritas.
In the one-zone SSC model, correlation between X-ray and VHE is expected, but it is not observed. However, the data is not sufficient to rule it out.

\thisfloatsetup{captionskip=-10pt}%

\thisfloatsetup{captionskip=-10pt}%

Figure~\subref*{fig:PG1553_nightly} shows the nightly \veritas and \xrt \lcs and the weekly \lat \lc for the \pg blazar.
Variability is seen in all energy bands, with X-ray featuring the most rapid changes in flux.
This is consistent with previous observations in VHE~\cite{Aleksic:2011rq, Abramowski:2015ixa} and studies of HE and X-ray \lcs~\cite{Ackermann:2015wda}.
The April 2012 flare~\cite{Abramowski:2015ixa} is seen in VHE and X-ray, but not in HE.
A stronger flare is seen in X-ray in December 2014, while no significant activity is observed in the \lat \lc. 
Unfortunately, no simultaneous \veritas data are available. 
Simultaneous \veritas and \xrt observations are available in the months following February 2015 however, where both energy bands feature a slowly decreasing flux.
Applying the \zdcf, no correlation is detected between the \veritas and \lat \lcs.
Evidence for correlation is observed between the \veritas and \xrt \lcs at a $3.8\sigma$ significance and with a lag of $-1^{+20}_{-14}$ days.
This result is consistent with a one-zone SSC model predicting a correlation between the synchrotron and the inverse-Compton emission as they relate to the same electron population.
The absence of correlation between VHE and HE was interpreted in Ref.~\cite{Abramowski:2015ixa} as a hint that the injected high energy particles predominantly emit VHE gamma rays. The HE flux may remain relatively unchanged if the flow of particles is not strong enough for their radiation during the acceleration or cooling phases to have an observable effect.

\thisfloatsetup{captionskip=-10pt}%
\begin{figure}[htp]
\begin{center}
\subfloat[]{\label{fig:PG1553_zdcf_fermi}\includegraphics[trim=0mm 0mm 0mm 0mm,clip,width=0.5\textwidth, page=1]{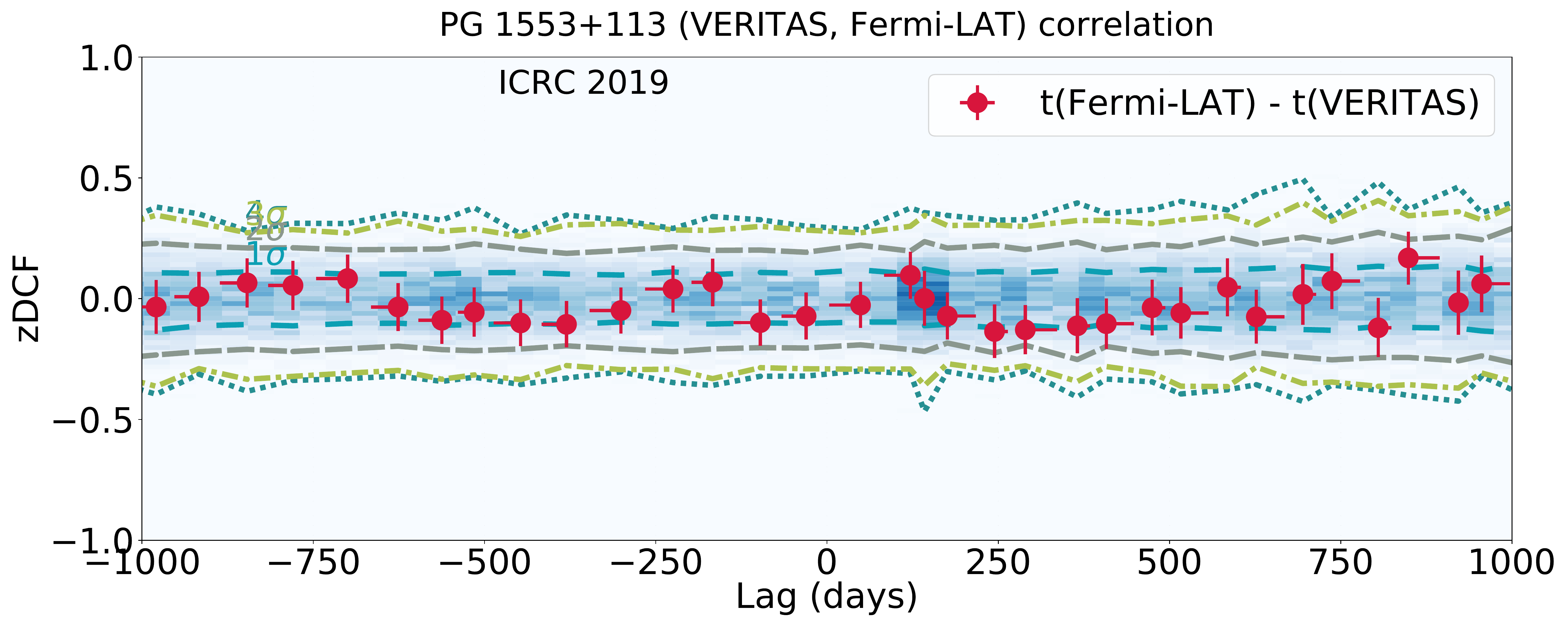}}
\subfloat[]{\label{fig:PG1553_zdcf_swift}\includegraphics[trim=0mm 0mm 0mm 0mm,clip,width=0.5\textwidth, page=1]{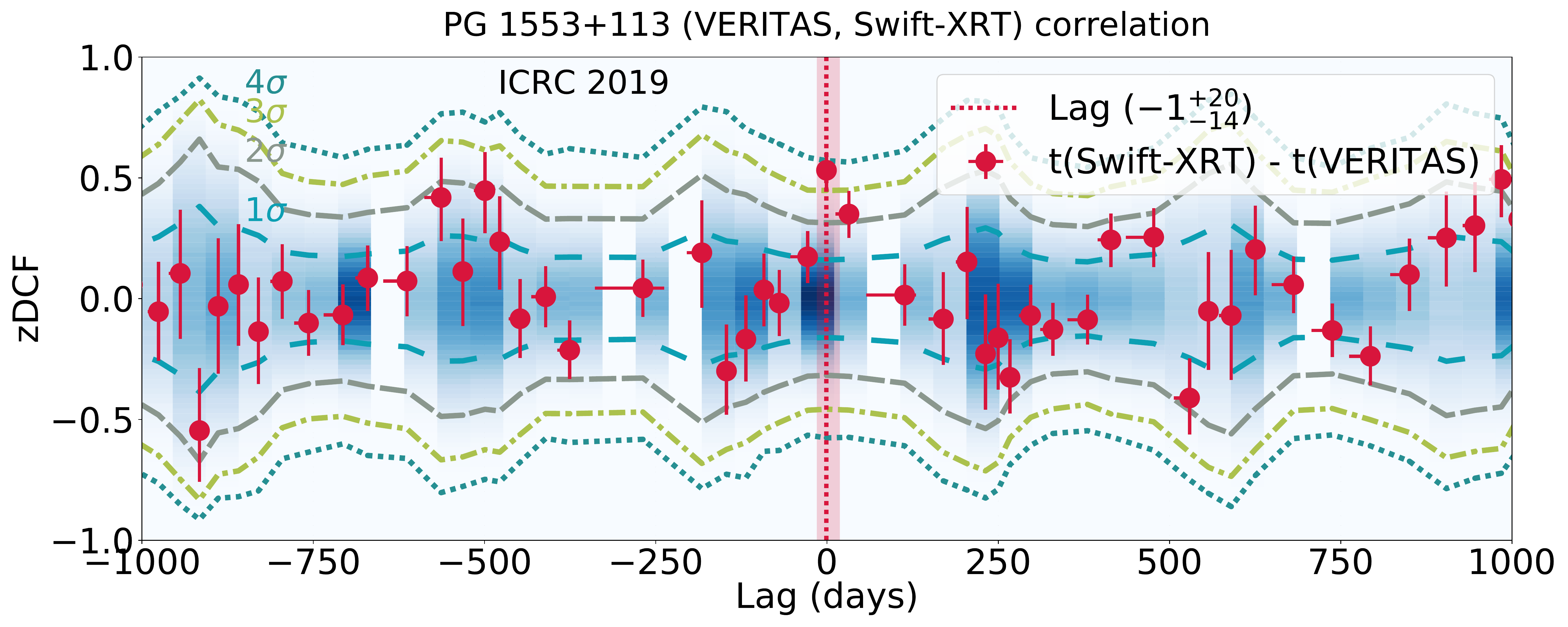}}
\caption{\label{fig:PG1553_zdcf_mwl}
The \zdcf between the \pg \lcs in \Subref{fig:PG1553_zdcf_fermi} VHE (\veritas) and HE (\lat); and \Subref{fig:PG1553_zdcf_swift} VHE (\veritas) and X-ray (\xrt). The vertical red dotted line and band indicate the most likely lag between the two bands and the corresponding 1-$\sigma$ confidence interval. The \zdcf distribution obtained from simulated \lcs is shown in the 2D histogram colour map (see text for details). The correlation significance levels correspond to the quantiles of the \zdcf distribution in each lag bin.
}
\end{center}
\end{figure}

\section{Secondary gamma rays contribution}

\noindent The gamma-ray flux from blazars is typically considered to originate from photons emitted at the source.
At high energies, these primary gamma rays are attenuated by their interactions with the extragalactic background light (EBL)~\cite{Salamon:1997ac}.
To overcome this attenuation, it was suggested that ultra-high-energy protons ($10^{17}$ -- $10^{19}$ \ev) produced at the source then proceed to interact with background photons and generate VHE secondary gamma rays along the line of sight~\cite{Essey:2009ju, Essey:2010er, Aharonian:2012fu}.
Such models predict that the observed VHE gamma-ray flux from blazars at redshift ${z > 0.15}$ and energy $E > 1 \tev$ contains a significant contribution from secondary gamma rays.
This contribution is expected to have distinct spectral and temporal features which should be visible in the observed gamma-ray flux. 
In particular, in the energy range dominated by secondary gamma rays, time variability on time scales shorter than $\sim$0.1 year should not be observed~\cite{Prosekin:2012ne}.
The detection of a contribution from secondary photons can help in setting upper limits on the magnetic fields along the line of sight, as the intergalactic magnetic fields must be $\lesssim 10^{-14}$~G for their contribution to be observable.

\begin{figure}[htp]
\begin{center}
\includegraphics[trim=0mm 0mm 0mm 0mm,clip,width=0.6\textwidth, page=1]{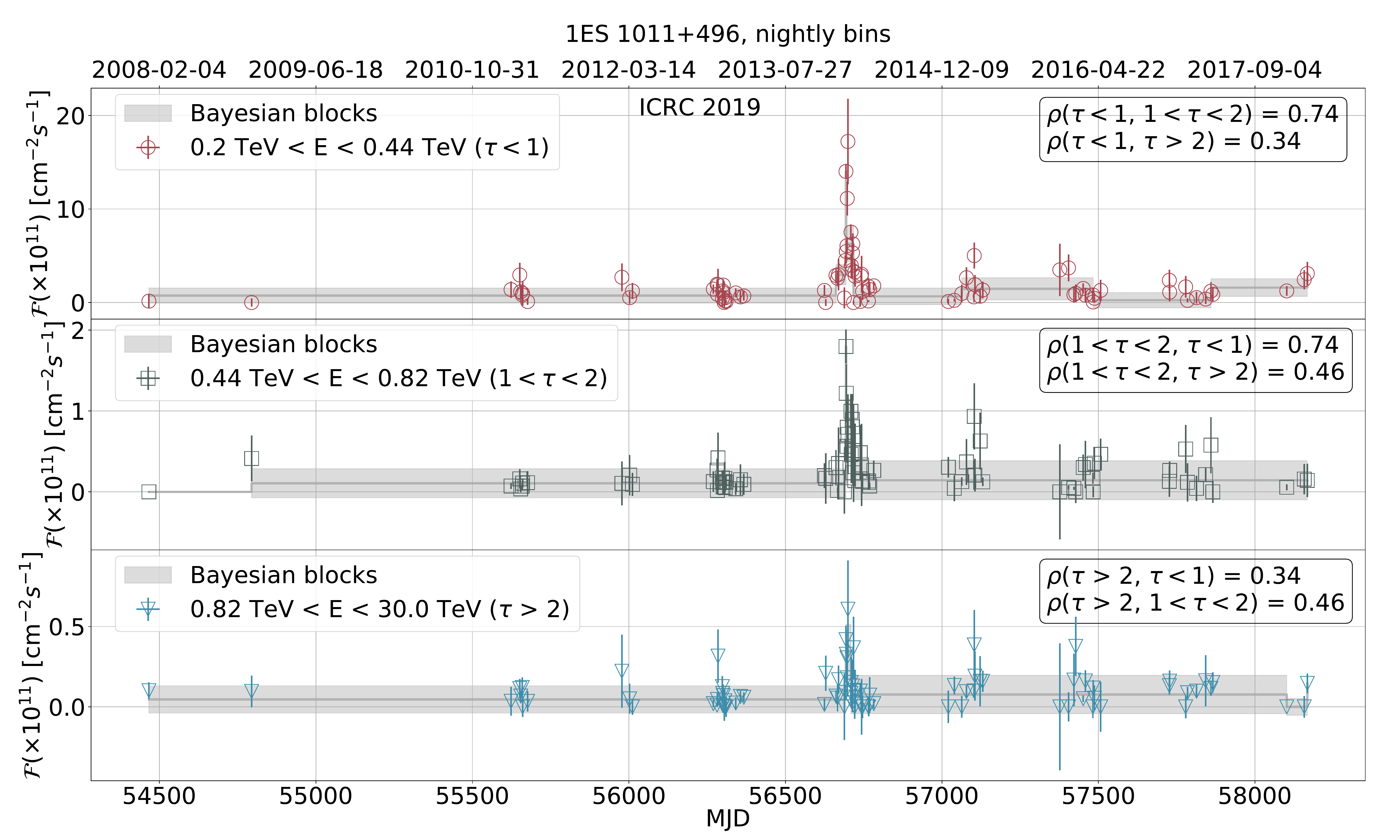}
\caption{\label{fig:1ES1011_vhe}
\veritas VHE \esTen \lcs binned nightly in three energy ranges, as indicated in the panels. The mean flux and the corresponding uncertainty in each Bayesian block are shown in grey. Pearson correlation coefficients between different bins are also displayed.
}
\end{center}
\end{figure}

\begin{figure}[htp]
\begin{center}
\subfloat[]{\label{fig:1ES1011_zdcf_tau2}\includegraphics[trim=0mm 0mm 0mm 0mm,clip,width=0.5\textwidth, page=1]{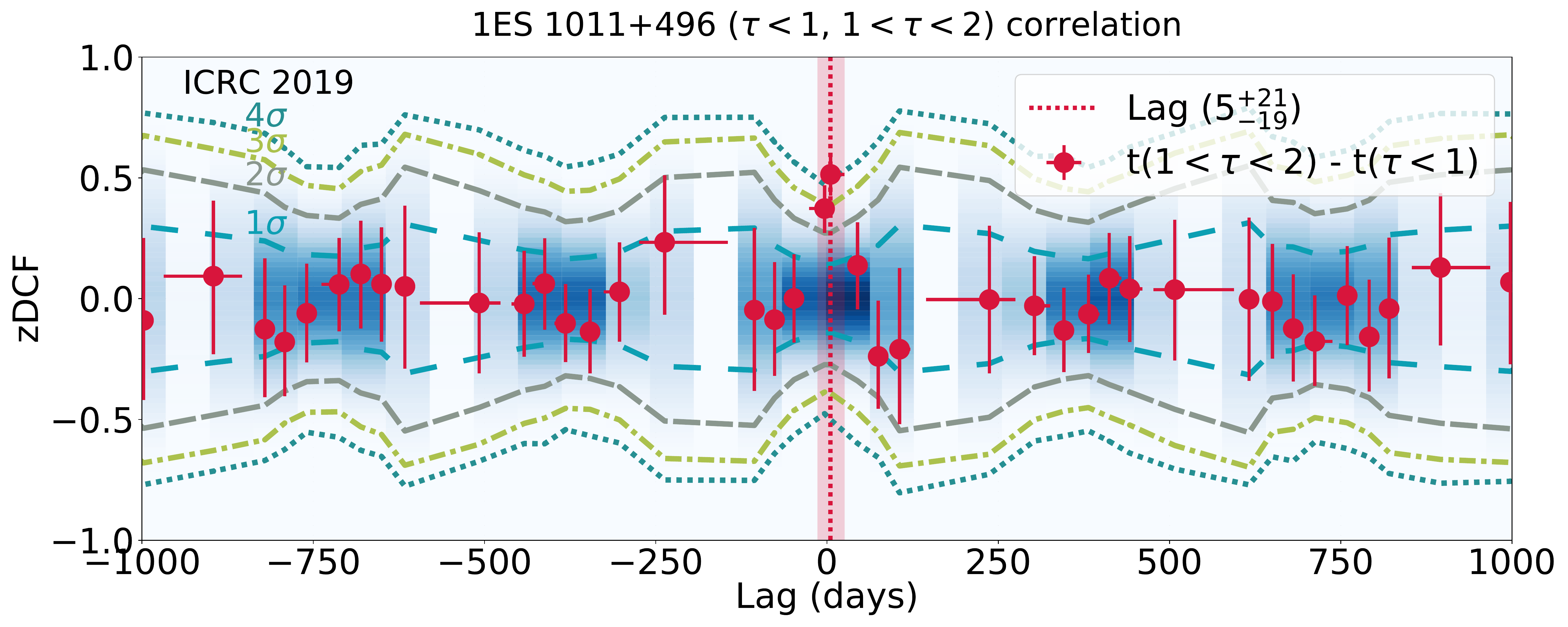}}
\subfloat[]{\label{fig:1ES1011_zdcf_tau3}\includegraphics[trim=0mm 0mm 0mm 0mm,clip,width=0.5\textwidth, page=1]{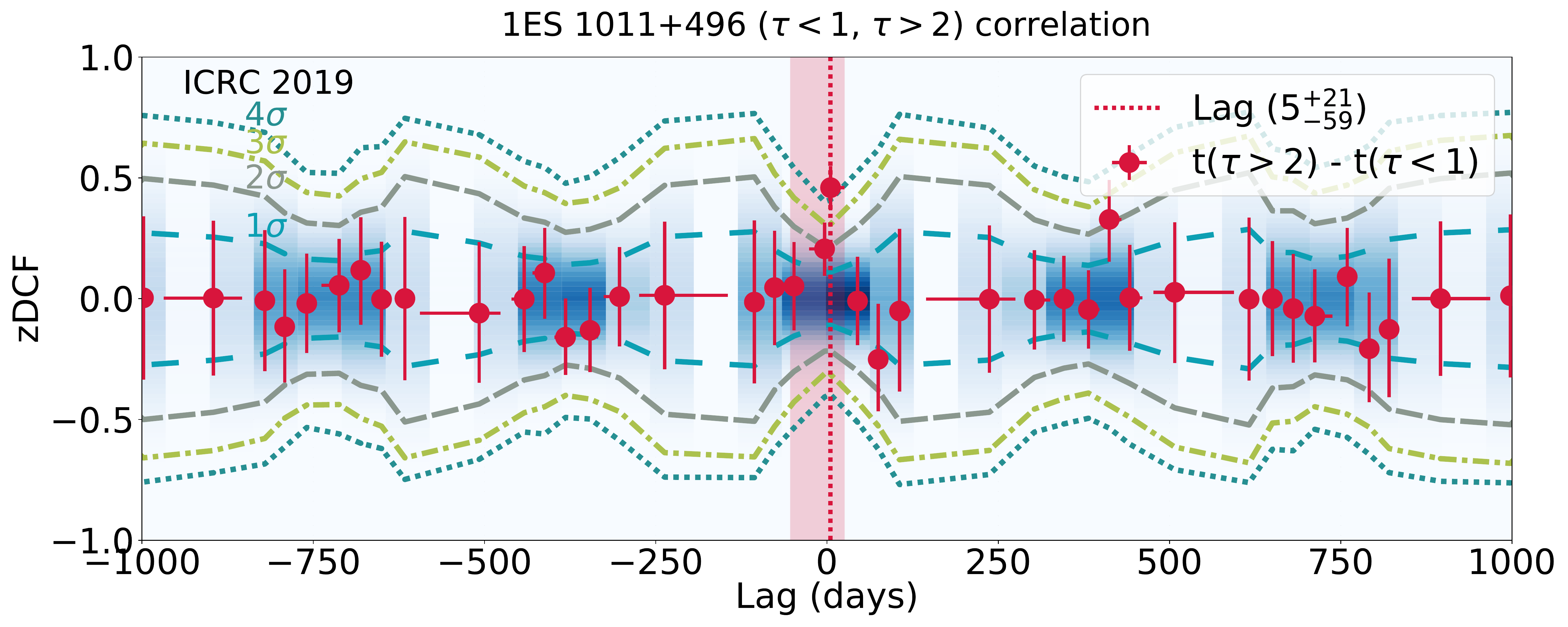}}
\caption{\label{fig:1ES1011_zdcf_vhe}
The \zdcf between the \esTen \lcs in the energy ranges \Subref{fig:1ES1011_zdcf_tau2} $\tau < 1$ and $1 < \tau < 2$; and \Subref{fig:1ES1011_zdcf_tau3} $\tau < 1$ and $\tau > 2$. The vertical red dotted line and band indicate the most likely lag between the two bands and the corresponding 1-$\sigma$ confidence interval. The \zdcf distribution obtained from simulated \lcs is shown in the 2D histogram colour map (see text for details). The correlation significance levels correspond to the quantiles of the \zdcf distribution in each lag bin.
}
\end{center}
\end{figure}

To test the contribution of secondary photons, the \veritas \lcs of the sources are divided into energy bins according to their opacity~\cite{Essey:2011wv}.
The EBL energy and redshift evolution in the Franceschini et al. 2008 model, $\tau(E, z)$,~\cite{Franceschini:2008tp} is used to set the energy ranges of three opacity bins, $\tau < 1$, $1 < \tau < 2$ and $\tau > 2$. 
These correspond to attenuation of the primary photons of $\mu \lesssim 63\%$, $63\% \lesssim \mu \lesssim 87\%$ and $\mu \gtrsim 87\%$ respectively.
The \lcs of \esTen in these opacity bins are shown in Figure~\ref{fig:1ES1011_vhe}. 
The February 2014 flare is visible and detected by the Bayesian block analysis in all three opacity bins.
Correlation between the \lcs is tested using the \zdcf method, as shown in Figure~\ref{fig:1ES1011_zdcf_vhe}.
A~$\sim 4\sigma$ correlation is observed between the $\tau < 1$ \lc and the other two \lcs, indicating a dominant contribution from the flare even for the $\tau > 2$ bin.

\thisfloatsetup{floatwidth=.6\textwidth,capbesidewidth=sidefil,
capposition=beside,capbesideposition={left,center}, postcode=afterfloat}

\thisfloatsetup{captionskip=-10pt, precode=beforefloat}%

A toy study is performed to assess whether this level of correlation can be observed within the secondary gamma rays hypothesis.
The latter predicts that primary photons emitted at the source contribute less than 13\% of the observed flux in the $\tau > 2$ bin.
The rest of the flux is assumed to come from secondary gamma rays.
Following this proposed composition, a simulated \lc is constructed to represent the $\tau > 2$ opacity bin, $\mathcal{L}^{\,\mathrm{sim}}_{\tau > 2} = 0.13 \times \mathcal{L}^{\,\tau < 1}_{\mathrm{primary}} + 0.87 \times \mathcal{L}^{\,\mathrm{sim}}_{\mathrm{secondary}}$.
The first element, $\mathcal{L}^{\,\tau < 1}_{\mathrm{primary}}$, is extracted from data corresponding to the $\tau < 1$ \lc, which are predominantly made up of primary photons.
The second component, $\mathcal{L}^{\,\mathrm{sim}}_{\mathrm{secondary}}$ is derived from $\mathcal{L}^{\,\tau < 1}_{\mathrm{primary}}$.
As the first step, the flare period is removed from $\mathcal{L}^{\,\tau < 1}_{\mathrm{primary}}$. 
This is done in order to assure that no fast variability is introduced to the secondary gamma ray contribution.
Next, the PSD of the resulting \lc is calculated, and used to simulate $\mathcal{L}^{\,\mathrm{sim}}_{\mathrm{secondary}}$.
Finally, the \zdcf between $\mathcal{L}^{\,\mathrm{sim}}_{\tau > 2}$ the $\tau < 1$ \lc is calculated. 
Correlation at the $\sim 3\sigma$ level is observed. 
This suggests that a 13\% contribution from primary photons during the flare is sufficient to produce the $\sim 4\sigma$ correlation observed between the $\tau < 1$ and $\tau > 2$ \esTen \lcs.
Therefore, based on the observed correlation, the hypothesis of ultra-high-energy protons producing secondary gamma rays along the line of sight cannot be ruled out.

\section{Conclusions}

\noindent The short- and long-term variability of xHBLs/HBLs was studied in the VHE, HE, and X-ray bands. 
Day-scale variability was observed in VHE during the \esTen flare, while a simultaneous increase in flux was seen in HE. 
Correlation at the level of $\sim 4 \sigma$ and with no significant lag was observed between the two \lcs, supporting the hypothesis that the same particle population emits gamma rays in those energy ranges.
The VHE and X-ray \lcs of \esTwelve present short time-scale variability with no apparent correlation between them.
Typically, this indicates that two different zones and/or particle populations are responsible for the VHE and X-ray emissions.
All three \pg \lcs feature variability, where correlation between VHE and X-ray is seen at $3.8\sigma$ level and no correlation is detected between those bands and the HE one.
The former is compatible with a SSC scenario in which the VHE and X-ray emission originate from the same zone.
The lack of correlation with the HE band joins the reported evidence for quasi-periodic modulation in the HE flux~\cite{Ackermann:2015wda} in challenging the one-zone SSC model. This tension prompts the need for more sophisticated models (e.g., Refs.~\cite{Caproni:2017nsh}~and~\cite{Tavani:2018lwu}).


The VHE \lcs were divided in three energy ranges, corresponding to three opacity bins.
A search for correlations between the resulting \lcs was performed in order to assess the contribution from secondary gamma rays produced along the line of sight from ultra-high-energy protons. 
The three \esTen \lcs were found to be correlated at a $\sim 4\sigma$ level, pointing to the existence of a significant contribution from primary gamma rays to all \lcs.
However, it was shown that such correlation could appear even in the case that primary photons contribute only 13\% to the $\tau > 2$ \lc.
The current dataset cannot therefore be used to set constraints on this model. 
For that purpose, a similar short-term variability and correlation needs to be observed with the $\tau > 3$ bin, in which the primary photons contribution is predicted to be $< 5\%$ and no correlation is expected.

\section*{Acknowledgments}
\noindent This research is supported by grants from the U.S. Department of Energy Office of Science, the U.S. National Science Foundation and the Smithsonian Institution, and by NSERC in Canada. This research used resources provided by the Open Science Grid, which is supported by the National Science Foundation and the U.S. Department of Energy's Office of Science, and resources of the National Energy Research Scientific Computing Center (NERSC), a U.S. Department of Energy Office of Science User Facility operated under Contract No. DE-AC02-05CH11231. We acknowledge the excellent work of the technical support staff at the Fred Lawrence Whipple Observatory and at the collaborating institutions in the construction and operation of the instrument.
This work made use of data supplied by the UK \textit{Swift} Science Data Centre at the University of Leicester.

\begin{multicols}{2}
\bibliographystyle{JHEP} 
\bibliography{xHBL_bib}
\end{multicols}

\end{document}